\documentclass{article}
\usepackage{iclr2021_workshop,times}

\usepackage{amsmath,amsfonts,bm}

\def\eqref#1{equation~\ref{#1}}

\def\1{\bm{1}}

\DeclareMathAlphabet{\mathsfit}{\encodingdefault}{\sfdefault}{m}{sl}
\SetMathAlphabet{\mathsfit}{bold}{\encodingdefault}{\sfdefault}{bx}{n}

\usepackage{graphicx}
\usepackage{hyperref}
\usepackage{url}
\usepackage{wrapfig}
\usepackage{subfigure}
\usepackage{caption}
\usepackage{multirow} 
\usepackage{etoolbox}
\AtBeginEnvironment{tabular}{\footnotesize}

\title{Recovering Barabási-Albert Parameters\\of Graphs through Disentanglement}

\author{Cristina Guzmán, Daphna Keidar, Tristan Meynier, Andreas Opedal \& Niklas Stoehr \\
Department of Computer Science\\
ETH Zurich\\
Zurich, Switzerland \\
\texttt{\{csolis,dkeidar,tmeynier,aopedal\}@ethz.ch} \&\\ \texttt{niklas.stoehr@inf.ethz.ch}\\
}

\def \hfillx {\hspace*{0.03\textwidth} \hfill}

\iclrfinalcopy 
\begin{document}

\maketitle





\begin{abstract}

Classical graph modeling approaches such as Erdős-Rényi (ER) random graphs or Barabási-Albert (BA) graphs, here referred to as stylized models, aim to reproduce properties of real-world graphs in an interpretable way. While useful, graph generation with 
stylized models requires domain knowledge and iterative trial and error simulation. Previous work by \cite{stoehr2019disentangling} addresses these issues by
learning the generation process from graph data, using a disentanglement-focused deep autoencoding framework, more specifically, a $\beta$-Variational Autoencoder ($\beta$-VAE). While they successfully recover the generative parameters of ER graphs through the model's latent variables, 
their model performs badly on sequentially generated graphs such as BA graphs,
due to their oversimplified decoder.
We focus on recovering the generative parameters of BA graphs
by replacing their $\beta$-VAE decoder with a sequential one. We first learn the generative BA parameters in a supervised fashion using a Graph Neural Network (GNN) and a Random Forest Regressor, by minimizing the squared loss between the true generative parameters and the latent variables. Next, we train a $\beta$-VAE model, combining the GNN encoder from the first stage with an LSTM-based decoder with a customized loss. 
\end{abstract}

\section{Introduction}
Knowing the generative procedure of a graph and its parameters allows for predicting, controlling and understanding its driving principles \citep{brugere2018network, barabasi2016network, newman2010networks}. One way to uncover the underlying generative parameters of such a procedure is through disentanglement -- an approach which has recently gathered a large interest in the research community. The aim is to find a representation of the data where the variables are disentangled in the sense that each variable corresponds to at most one generative parameter. Disentangled representations have been successfully found for image data  \citep{burgess2019monet, steenkiste2019abstract, leeb2020structural, besserve2019counterfactuals}. However, disentanglement of graph data is largely unexplored, with a few exceptions \citep{Ma_DisGCN, Guo_2020, stoehr2019disentangling}. 
\\\\
Previous work \citep{stoehr2019disentangling} has proposed a disentanglement framework with which they recover the generative parameters of graphs generated by stylized graph models. Stylized graph generation models are graph generation algorithms
that typically follow a fixed procedure in order to capture certain characteristics of real-world graphs. While useful, these models 
require pre-specified parameters. Furthermore, finding the appropriate values of these parameters typically demands either domain knowledge or time-consuming simulations. With a disentanglement framework however, one can find these generative parameters unsupervised from a given dataset.
\\\\
A commonly used stylized model is the Erdős-Rényi (ER) random graph generator, which defines the number of nodes $n$ and the uniform linking probability $p$ between any two nodes in the graph. In this generation process the individual edges are sampled independently of each other and the graph can thus be generated accurately "one-shot", meaning that the procedure can be non-iterative. However, most real-world graph datasets are generated through a sequential process. Therefore, the ER model does not capture many of the properties typically observed in real-world graphs, such as the tendency of nodes to link to more connected nodes, a phenomenon known as preferential attachment. Graphs that exhibit this property are called scale-free, and they evolve dynamically, continuously growing with the addition of new nodes. The Barabási-Albert (BA) Preferential Attachment model \citep{BarabasiAlbert} is a stylized model that was devised in order to more accurately model such graphs, in which edges are generated sequentially. As the properties of scale-free graphs are inherently the result of an iterative process, we are not aware of a method for generating them "one-shot". The $\beta$-Variational Autoencoder ($\beta$-VAE) architecture \citep{higgins2017beta} suggested by \cite{stoehr2019disentangling} is capable of both generating graphs and recovering compact, disentangled representations of their generative parameters. While their approach performs well when applied to ER graphs, their decoder generates the graphs one-shot 
which hinders their performance on the scale-free BA graphs. 

Focusing on data generated from the non-linear extension of BA \citep{nonlinearBA},
the main contribution of this work is creating a graph VAE with a sequential decoder, for finding compact graph representations corresponding to the generative parameters of BA graphs. In this paper we first show that it is possible to recover the generative parameters, by learning them in a supervised fashion using a Graph Neural Network (GNN) and a Random Forest (RF) model, using the mean squared error (MSE) loss between the true generative parameters and the predictions. Next, we train a $\beta$-VAE model, combining the GNN encoder from the previous stage with an LSTM-based decoder, in order to obtain a disentangled representation of the generative parameters. Creating a sequential graph decoder network is especially challenging, as it should mimic a sequential random generation process, with random sampling at each step. Such a process is not naturally differentiable with respect to the latent variables, and typically requires tricks to enable backpropagation. As such, we adapt the LSTM model to the graph setting for this task.

\section{Models and Methods}
\label{sec:methodology}
We use a synthetic dataset that is generated with the BA model. 
The modeling approach is split into two stages: parameter prediction via supervised training and disentanglement via unsupervised training. In the first part we predict the generative factors of the BA graphs using both a Random Forest and a GNN. 
In the second part, we implement a $\beta$-VAE with a sequential LSTM
decoder network, to find disentangled latent representations of the generative parameters of the BA graphs.

\paragraph{BA Graph Generation} The non-linear extension of the BA model is parameterized by the number of nodes $n$, an edge generation integer $m$ and the exponential of the node degree $\alpha$. In the standard BA model \citep{BarabasiAlbert}, we set $\alpha=1$ and get a linear dependency on the node degrees. In real-world scale-free graphs, preferential attachment can also be non-linearly dependent on the node degrees and the extended BA model \citep{nonlinearBA} accounts for these scenarios through the parameter $\alpha$. The description of the generative process can be found in Appendix \ref{sec:appendix-ba} and illustrations of BA graphs with different values of $\alpha$ are shown in Figure \ref{fig: BAplots}.






\subsection{Predicting Generative Parameters}
In this stage, we learn the generative parameters in a supervised manner. We train a model to output predictions for the generative parameters, and take the MSE between the model's output and the true generative parameters to be our training loss. 
As a first step, we want to verify that the generative parameters are learnable from the data. To do so, we extract feature vectors from the input graphs, and then train a Random Forest (RF) Regressor to predict the parameters from these vectors. Note that this step is only for the purpose of verification, as the RF model does not allow for backpropagation and can thus not be included in an autoencoder framework. 
Furthermore, a non-neural model requires manual feature extraction, whereas a GNN can automatically extract features from the graphs. 
\\\\  
As such, we proceed to implement a GNN to predict the generative parameters, being adapted from \cite{stoehr2019disentangling}. 
The outputs of the GNN can be interpreted as the latent variables in an autoencoder model, with a BA generator as the decoder. The BA generator is non-differentiable and lacks learnable parameters however, which prevents unsupervised training via minimizing the reconstruction loss. A high-level architecture can be seen in the top part of Figure \ref{fig:overview}.

\subsection{Generative Parameters as Disentangled Latent Variables}
\begin{figure}[t!]
    \centering
    \includegraphics[width=0.78\columnwidth]{
    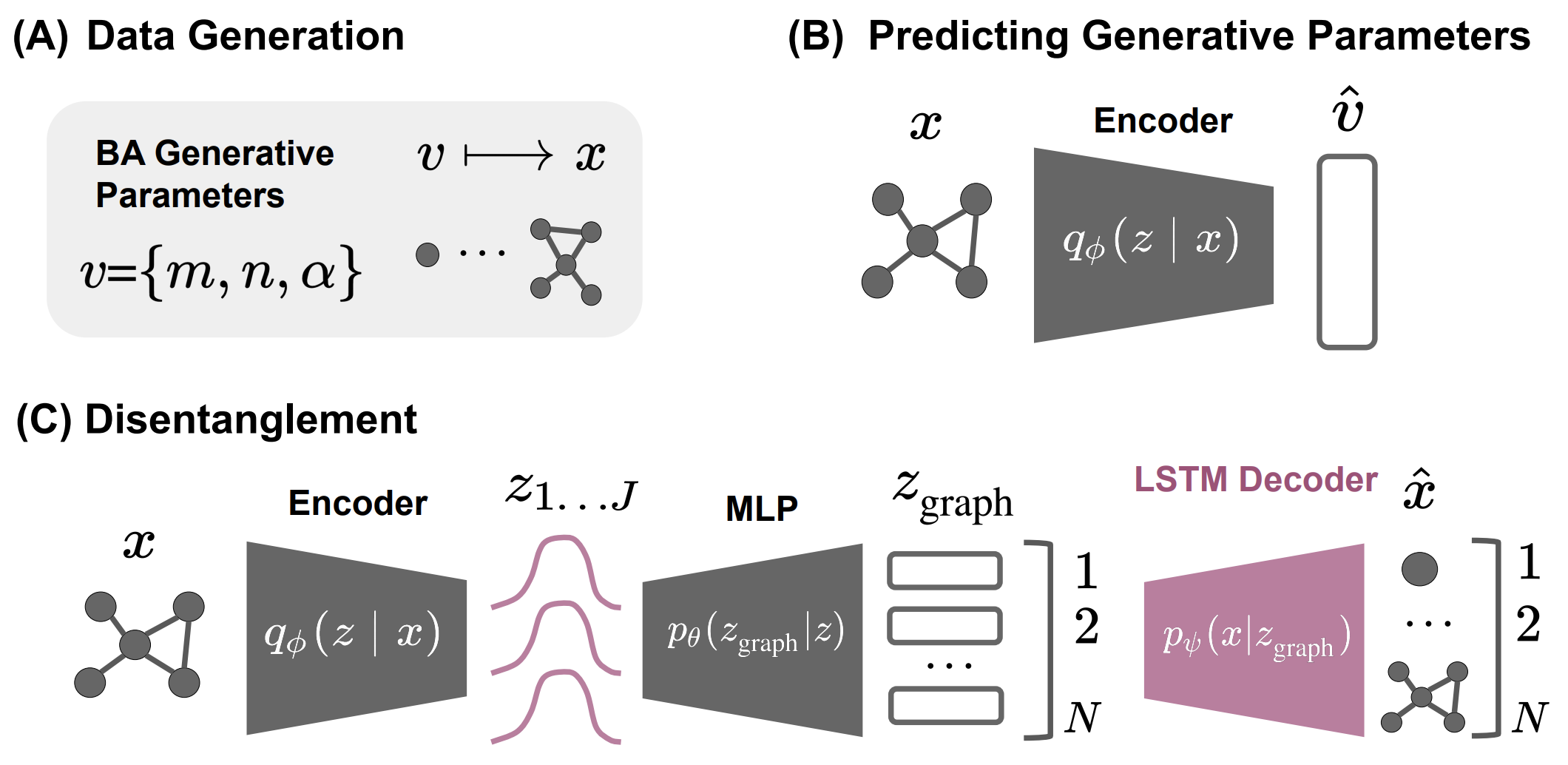}
    \caption{(A) Data generation (B) Predicting generative parameters in a supervised setting (C) Disentangling latent variables to recover generative parameters in an unsupervised setting}
    \label{fig:overview} 
\end{figure}

To obtain a VAE, we keep the GNN encoder used in the prediction stage, and implement a differentiable decoder in the form of an LSTM. The latent variables produced by the encoder are passed through a deconvolutional layer to obtain $N$ graph representations, $N$ being the number of generations steps. 
 These are then passed to the LSTM decoder. The model is trained with the standard $\beta$-VAE objective, composed of a scaled KL-divergence term enforcing the variational posterior to be similar to an isotropic Gaussian, and a reconstruction term comparing the input and output graphs. We use a customized reconstruction term to incorporate a set of constraints to adhere to a graph generation process; namely enforcing symmetry, no self-connections, increasing number of nodes and edges in each step, and starting with an empty graph. This is added in order to replicate the generation process of a sequential stylized model. A high-level architecture of this stage can be seen in the bottom part of Figure \ref{fig:overview}.
 
 \paragraph{Mutual Information Gap}
Among the many metrics for assessing the quality of disentangled latent representations\footnote{See \cite{locatello2019fairness} and \cite{steenkiste2019abstract} for a comparison of different disentanglement metrics in fairness and abstract visual reasoning respectively.}, we choose MIG as it captures axis-alignment and is unbiased \citep{chen2018isolating}.
The MIG of latent variables $z_{1,...,J}$ with respect to generative factors $v_{1,...,K}$ is defined as:
\[\frac{1}{K}\sum_{k=1}^K\frac{1}{H(v_k)}(I(z_{j^{(k)}};v_k)-\max_{j\neq j^{(k)}}I(z_j;v_k))\] where $I$ is the mutual information, $H$ is the entropy and $j^{(k)}=\arg\max_jI(z_j;v_k)$. Intuitively, the MIG averages the difference between the largest mutual information and the second largest mutual information for each generative parameter with respect to the latent variables. It is then normalized by the entropy to be between $0$ and $1$, where a larger MIG value represents better disentanglement.




\section{Results} 
\label{sec:results}
We assess the performance of the supervised GNN model in predicting the generative parameters, for both linear and non-linear BA graphs. The training and test sets are of size $16,000$ and $4,000$ respectively. The number of nodes $n$ is fixed to be $50$, $m$ is sampled as an integer uniformly at random between $1$ and $49$, and for the non-linear case $\alpha$ is sampled uniformly at random between $1/3$ and $3$ in order to capture graphs of both sub-linear and super-linear degree distribution\footnote{Further implementation details can be found at \\ https://github.com/AndreasOpedal/ba-graph-disentanglement}.
The results of the two learning tasks are shown in Table \ref{table: vanilla} together with the results from the RF validation model. 

 The GNN learns the parameters of the linear BA model with a satisfactory performance, but does not manage to learn $\alpha$, while the RF model manages to learn $\alpha$ with a significantly lower MSE and a Pearson correlation coefficient of 0.89 between the true and predicted $\alpha$, indicating that $\alpha$ is in fact learnable from the graph data. 
The predicted parameter values can be used to generate new graphs, using the BA graph generator. These can be compared to the original graphs for performance assessment, see Appendix \ref{sec:appendix-hist} for more details. 
For the LSTM-based $\beta$-VAE, we compute the MIG score between the generative parameters and the latent variables. The model is trained on two BA datasets, a linear one with $\alpha=1$, and a non-linear one with $\alpha \in (\frac{1}{3},3)$. Both datasets consist of $200$ graphs, with node sizes varying from $3$ to $12$, and the model is trained for $200$ epochs.
The MIG scores are presented 
in Table \ref{tab: MIG}. We see that when trained on linear BA data the model displays the best performance, as expected since this is an easier task. 
We further note that the parameters $n$ and $m$ in the BA model are not independent. Since independence is enforced by the KL-term when training a $\beta$-VAE we also examine the MIG with respect to the parameter $\frac{m}{n}$ instead of $m$. This parameter is independent from $n$, but together with $n$ completely describes $m$. We observe in Table \ref{tab: MIG} that the MIG scores for $\frac{m}{n}$ is indeed higher.


\begin{table}[t!]
        \begin{minipage}{0.5\textwidth}
            \centering
           \begin{tabular}{|c|c|c|c|}
\hline
\multirow{2}{*}{\textbf{Task \& Data}}&
\multicolumn{3}{c|}{\textbf{MSE on test set}} \\ \cline{2-4}
 & $\mathbf{n}$ & $\mathbf{m}$ & $\mathbf{\alpha}$ \\ \hline
GNN \&  Linear BA  & 0.0005  & 0.128 & - \\
GNN \&  $\alpha\sim\mathcal{U}(\frac{1}{3},3)$      
&  0.005 & 0.257 &  0.590 \\ 
Random Forest \& $\alpha\sim\mathcal{U}(0,3)$ & 0.000 & 1.823 & 0.194 \\\hline
\end{tabular}
            \caption{Model performance in supervised setting.}
            \label{table: vanilla}
        \end{minipage}
        \hfillx
        \begin{minipage}{0.4\textwidth}
            \centering
           \begin{tabular}{|c|c|c|}\hline
        \textbf{Data} &\textbf{MIG} \\
        \hline
        Linear BA & 0.26\\
        Linear BA ($m/n$) & 0.29 \\
        Non-linear BA & 0.12\\\hline
    \end{tabular}
    \caption{MIG scores in unsupervised setting.}
    \label{tab: MIG}    
        \end{minipage}
\end{table}
\section{Discussion}
\label{sec:discussion}
\paragraph{Encoder Architectures} In the prediction setting, we achieve satisfactory results for learning the parameters with a simple Random Forest model that uses manually extracted features as input. We do not manage to learn $\alpha$ with the GNN, which also contributes to poor results in the disentanglement stage, where we use this architecture as the encoder. However, the performance of the RF model attests that the graph data holds sufficient information about the $\alpha$ parameter for a machine learning model to learn it. We thus believe that other GNN architectures should be experimented with to obtain better performance on this task. In particular, the graph convolutional layer used in our GNN requires adjacency matrices as input and output and as such depends on a fixed number of nodes of the graph\footnote{For the experiments varying $n$ the work-around used was to zero-pad the elements of adjacency matrices of some specified maximum size.}. There exist various graph convolutional layers that are invariant to the number of nodes, outputting embeddings on node level instead of graph level \citep{Bacciu_2020}, that could be used in future work.

\paragraph{Categorical Latent Variables}
We use Gaussian distributions as priors on the $\beta$-VAE's latent variables. However, both $n$ and $m$ are categorical variables, and $\alpha$ is bounded to a fixed-length interval, so Gaussians do not provide an optimal fit. In the linear BA case, we could sample from a discrete distribution, for instance using the Gumbel-Softmax trick \citep{jang2017categorical}. 
Implementing this in the non-linear case is not as straightforward, since $\alpha$ is not discrete.

\paragraph{Generation Model} A major challenge is the backpropagation through the decoder model, which essentially requires a graph generation process that both sequentially generates a graph and is differentiable with respect to the latent variables. The LSTM model can be considered sequential as it takes a graph representation as input in every time step. However, it conditions on the full generation history which is expensive and not necessary in order to replicate a BA graph generation process, in which each generation step conditions only on the previous state of the graph. Thus, a model with a first-order Markov assumption would have been sufficient. In real-world graphs however, the generation process might depend on earlier states of the graph as well, motivating our choice of a model with long-term dependency. We leave it for future work to discover the dynamics and dependencies of real-world graph generation. Other drawbacks of the LSTM model are long training times as well as it not being a natural fit for graph data, requiring tricks to output an adjacency matrix. We note the existence of sequential GNN architectures \citep{li2018learning, liu2018constrained}. However, the former conditions on the full generation history and the latter is tailored specifically towards molecule generation.
\paragraph{Node Attributes} In this work, we focus on graphs without node attributes. However, the same disentanglement methods can also be applied to graphs where the nodes come with attributes -- a feature present in most real-world graphs.
An interesting future direction could be to identify and disentangle the dependency of node attributes on the graph topology.
\section{Conclusion}
\label{sec:conclusion}
We recovered the generative parameters of the Barabási-Albert model and its non-linear extension through disentanglement. 
To this end, we emphasized the differences between sequential and non-sequential graph generators, and the scarcity of interpretable neural graph generators adhering to a sequential generation process. We hope to point out a promising and fascinating research direction with plenty of room for future improvement.
\bibliography{refs}
\bibliographystyle{iclr2021_workshop}
\newpage
\appendix
\section{Appendix}
\label{sec:appendix}
\subsection{Barabási-Albert generation procedure}
\label{sec:appendix-ba}
The non-linear extension of BA \citep{nonlinearBA} is parameterized by the number of nodes $n$, an edge generation integer $m$ and the exponential of the node degree $\alpha$. The generative process is defined as follows:\\
\begin{enumerate}
    \item Initialize a graph with $m$ nodes.
    \item At each step, until graph has $n$ nodes, add a new node and connect it to $m$ nodes randomly sampled from the following categorical distribution: 
    \begin{equation}
        \label{equation: BA}
        p(k_i)=\frac{k_i^{\alpha}}{\sum_{j}k_j^{\alpha}}
    \end{equation}
    where $k_i$ is the degree of node $i$, and  $p(k_i)$ is the probability that the new node connects to node $i$.
\end{enumerate}

\begin{figure}[!ht]
    \centering
    \includegraphics[width=0.7\columnwidth]{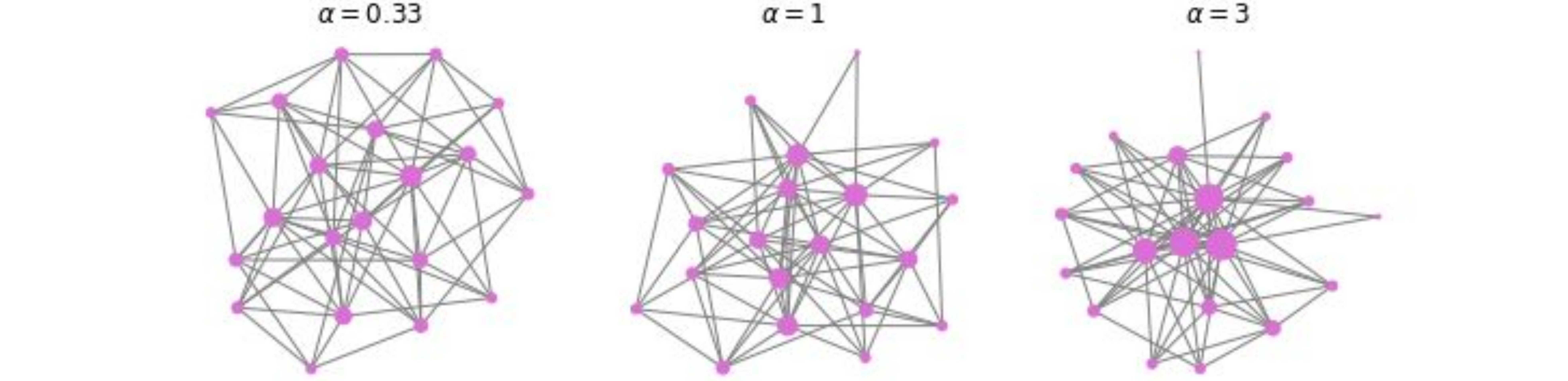}
    \caption{BA graphs with $20$ nodes and $m=5$ under different values of $\alpha$. \textbf{Left}: in the sub-linear case, graphs are similar to random ER graphs, with no presence of hubs. \textbf{Center}: BA model. \textbf{Right}: Nodes with higher degree are more attractive, leading to fewer but larger hubs. Nodes with higher degrees are displayed as larger.}
    \label{fig: BAplots} 
\end{figure}

\subsection{Performance Assessment of Parameter Prediction}
\label{sec:appendix-hist}
\begin{figure}[!ht]
    \centering
    \includegraphics[scale=0.34]{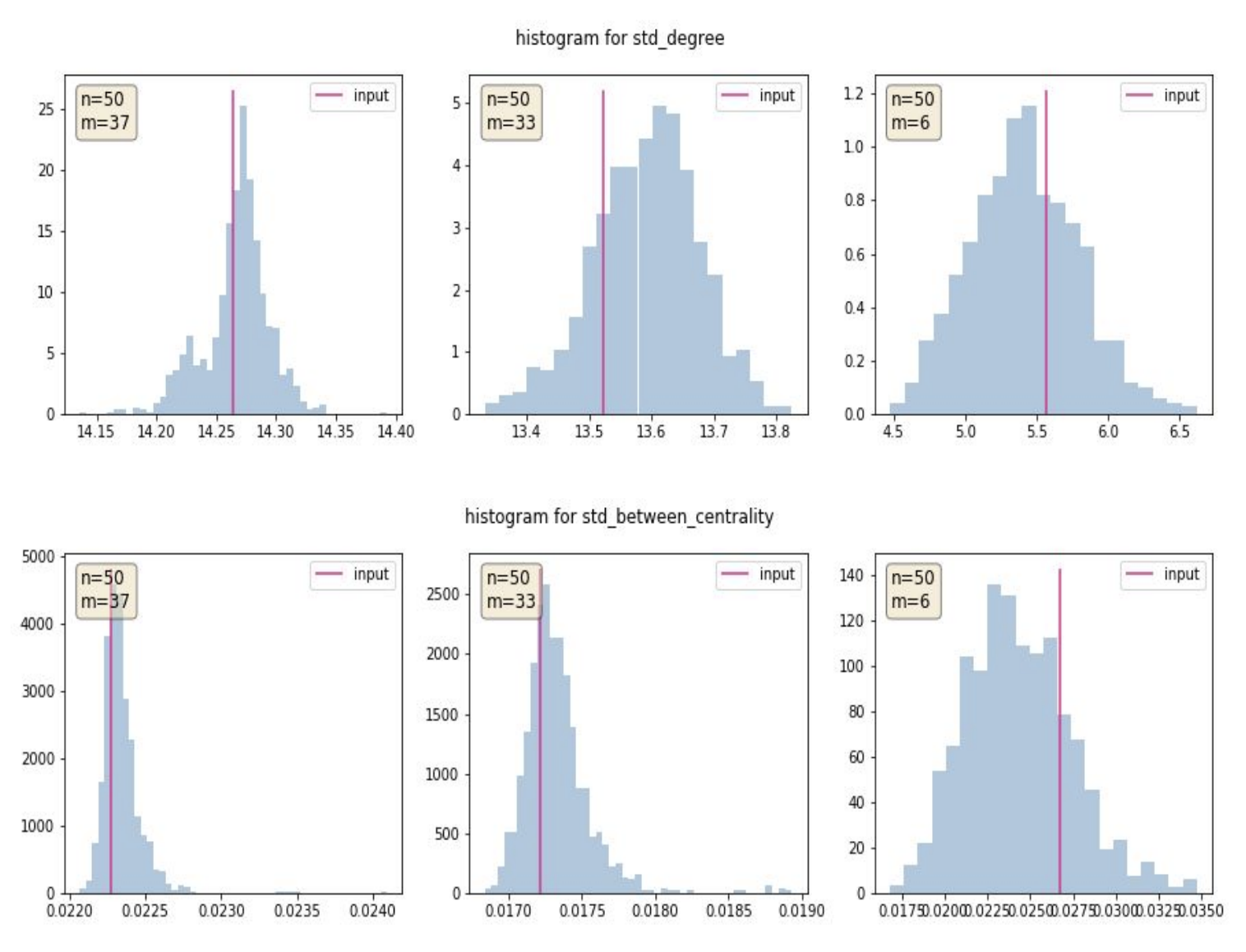}
    \captionsetup{justification=centering}
    \caption{Graph measures of test graphs compared to $1,000$ generated graphs from linear BA model, with true generative parameters of the test graphs.}
    \label{fig: standard hist}
\end{figure}

\begin{figure}[!ht]
    \centering
    \includegraphics[scale=0.34]{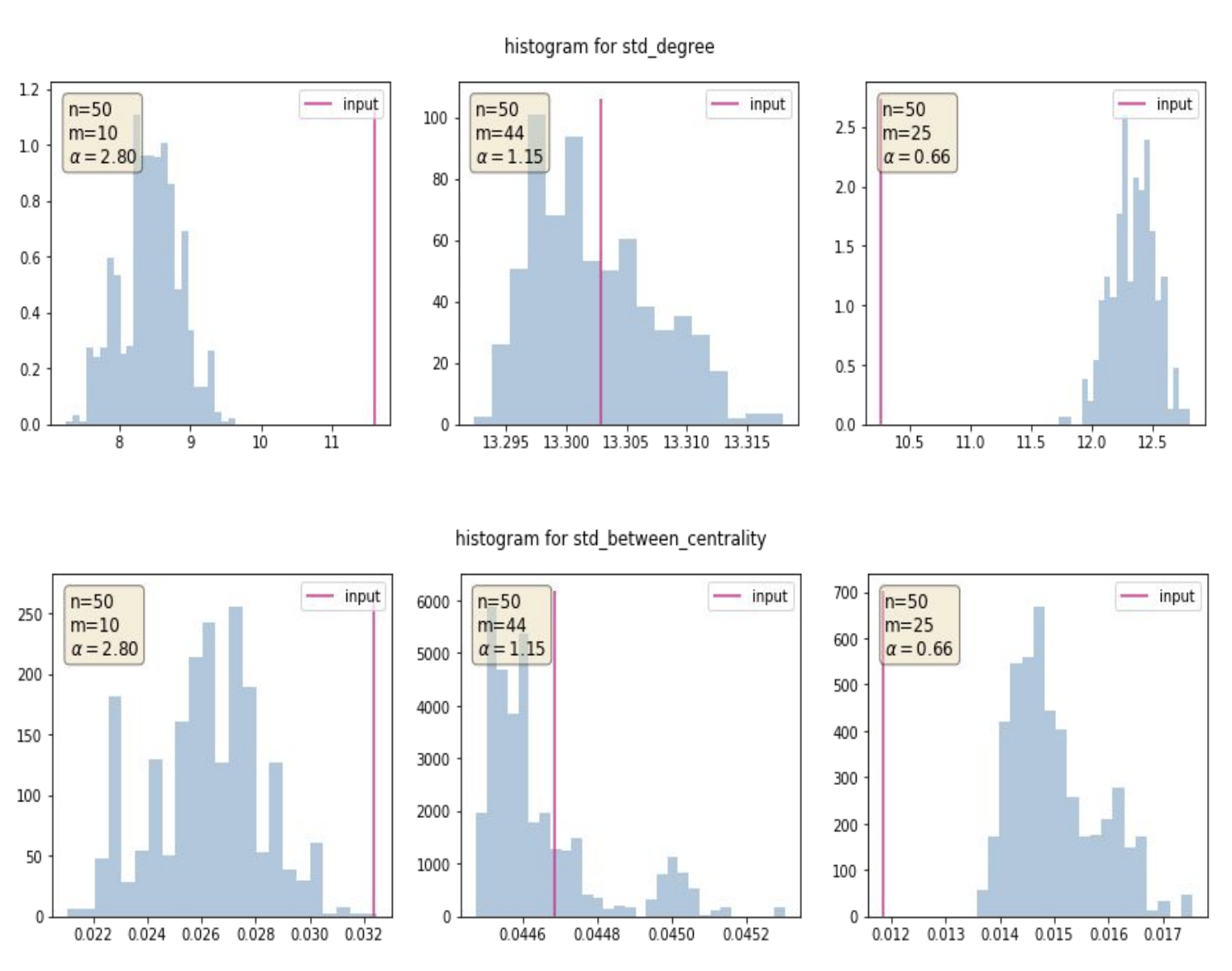}
    \captionsetup{justification=centering}
    \caption{Graph measures of test graphs compared to $1,000$ generated graphs from non-linear BA extension model, with true generative parameters of the test graphs.}
    \label{fig: non-linear hist}
\end{figure}

For each test graph, we compute the standard deviation of degree and betweenness centrality for $1,000$ generated graphs and compare to those of the input graph. We consider the model's predictions of parameters reasonable if the input graph properties fall within the distribution given by the generated graphs. The plots for a sample of three test graphs from the linear \& non-linear BA models are shown in Figure \ref{fig: standard hist} \& \ref{fig: non-linear hist} respectively. As seen in these examples, the measures of graphs generated from the linear BA model more closely follow their induced graphs' distribution.


\end{document}